\documentstyle[12pt]{article}
\begin{document}
\title{Nonlinear Maxwell Theory and Electrons in Two Dimensions.}
\author{Artur Sowa\\
        Department of Mathematics,
        Yale University\\
        10 Hillhouse Avenue,
        New Haven, CT 06520}
\date{ }
\maketitle
\newtheorem{th}{Theorem}
\renewcommand{\l}{\bigtriangleup_{A}}

\begin{abstract}

We consider a system of nonlinear equations that extends the Maxwell theory.
It was pointed out in a previous paper that symmetric solutions of these 
equations display properties characteristic of magnetic oscillations. 
In this paper I study a discrete model of the equations in two dimensions. 
This leads to the discovery of a new mechanism of vortex lattice formation. 
Namely, when a parameter corresponding to a magnetic field normal to the 
surface increases above a certain critical level, 
the trivial uniform-magnetic-field solution becomes, in a certain sense,
unstable and a periodic vortex lattice solution emerges. 
The discrete vortex solutions are proven to exist,
and can also be found numerically with high accuracy. 
Description of  magnetic vortices given by the equations is
optical in spirit, and may be particularly attractive in the context of 
high-$T_c$ superconductivity and the quantum Hall effects. 
Moreover, analysis of parameters 
involved in the discrete theory suggests existence of continuous domain 
solutions---a conjecture that seems unobvious on grounds of the current 
topological and variational methods.   
\end{abstract}

\newpage
\vspace{1 cm} \vbox spread 3in{}
\includegraphics{gls_im.ps}
\newpage
\section{The proposed physical model}

Let $A$ be the electromagnetic vector potential, and let 
the corresponding electromagnetic field be denoted by $ F_A = dA$. 
Further, let $f$ be a real valued function.
Consider the following system of equations
\begin{equation}
 dF_{A}=0
\label{syst0}
\end{equation}

\begin{equation}
 \delta (fF_{A})=0
\label{syst1}
\end{equation}

\begin{equation}
 -\bigtriangleup f +|F_{A}|^{2}f=\nu f.
\label{syst2}
\end{equation}
The main goal of this paper is to describe a vortex-lattice type 
solution of this system of equations in two dimensions (cf. figure)
by means of a finite difference approach. Results obtained in this way
suggest that in the continuous domain limit such a solution
consists of a continuous function $f$ and a connection $A$ whose
curvature $2$-form $F_A$ is also continuous. Moreover, the equations 
are satisfied in the classical sense almost everywhere. 
Note that the displayed numerical solutions are not defined at the 
vortex points. (Here, the horizontal axes are indexed by 
discrete lattice points.) Moreover, their scaling should be matched 
to a given physical system.

As explained in \cite{sow1} (and in a simplified Abelian 
version in \cite{sow2}), the particular form of this system 
is suggested by the geometry of principal $U(1)$-bundles.
Purely phenomenologically (\ref{syst0})-(\ref{syst2})
can be motivated by evoking heuristics used
frequently in the optical literature. 
Namely, it is often assumed in optics that the nonlinearities arising in the 
interaction of radiation with matter can be accounted for  
by representing $f$ as a series in tensorial powers of $F_A$ with coefficients
characteristic of the material.
Subsequently, one attempts to deduce properties of the coefficients
from a microscopic theory. The shift of paradigm in this paper
consists in assuming that $f$ depends in a geometrically invariant 
way on $F_A$ via equation (\ref{syst2}),
and $f$ remains essentially independent of the material except for a 
simple scaling. 
We will see below that in contrast
to the soft optical nonlinearities, (\ref{syst0})-(\ref{syst2}) 
cannot be understood as a small perturbation of a linear system.  
However, interaction of radiation with matter is described from the ``point 
of view'' of the former,
and the electronic processes inside  matter are never discussed directly.

In order to explain superconductivity by means of a microscopic theory,
one needs to display a mechanism that will let fermions 
overcome the obstacle imposed by the Pauli exclusion principle.
A solution given by the famous BCS theory
is based on the observation that when thermic noise is sufficiently low, it is 
energetically favored for electrons to join in pairs, 
known as Cooper pairs, which behave like bosons. The BCS theory is in
a certain correspondence to the Ginzburg-Landau equations.   
These are nonlinear equations for a complex valued function $\psi$, 
often interpreted as an order parameter governing  collective behavior of Cooper pairs. 
Periodic solutions of these equations in the form of vortices 
have been found by Abrikosov in \cite{abr}.
It must be emphasized that strict validity of the BCS/GL approach,
at least in its classical s-wave pairing version, is
limited to low temperature superconductivity of metals.

On the other hand, type II superconductivity is 
known to occur in materials structurally different from metals, 
like YBCO, and at relatively high critical temperatures. 
As many researchers pointed out, this suggests that mechanisms other 
than those encompassed by the BCS theory 
may be responsible for high temperature superconductivity. 
Those aspects of solid state theory that go beyond BCS 
seem particularly attractive in terms of the possibility of merging
with the nonlinear Maxwell equations. 
It is possible that the new 
mathematical pattern introduced in this paper will be helpful in the 
description of the interaction of magnetic fields with composite 
particles. For illustration, consider the proposition that $f$ describes a locally
varying {\em filling factor}. 
In this interpretation, a part of the field gets entrapped in composite bosons, 
composite fermions, and Laughlin quasiparticles, which in turn  
see only the remaining $\frac{1}{f}$-fraction of the field.
If $f$ is a constant, this allows one to replace an electron picture
with a composite particle picture, a suggestion that was present in science
already ten years ago. 
However, if $f$ is a vortex-type solution, the composite particles will 
feel the vortex in $F_A$, which should induce Josephson-type effects.
Thus, while microscopic theory is always 
constructed with an {\em a priori} fixed filling factor, 
(\ref{syst0})-(\ref{syst2}) would reflect the behavior of magnetic 
fields on a coarser scale.

The above is meant to evoke some of the basic 
notions and ideas present in modern materials science. 
More thorough reviews can be found in the articles featured in the very 
incomplete list of references below (\cite{andr1}-\cite{zhang}).

\section{Mathematics of a finite difference approximation}
 
Consider the system (\ref{syst0})-(\ref{syst2}) 
on a two-dimensional flat torus $T^2$.
In this case $\delta (fF)=0$ implies $\star F=\frac{B}{f}$ for
a constant $B$. In addition, $dF=0$
and $F$ is the curvature of a certain connection $A$, provided its cohomology class
satisfies $[F]\in 2\pi Z$.
Thus, the system of equations reduces to
\begin{equation}
 -\bigtriangleup f +\frac{B^2}{f}=\nu f ,
\label{tor1}
\end{equation}
and
\begin{equation}
\int_{T^2}\frac{B}{f}dV = 2\pi K \qquad\mbox{for an integer}\quad K.
\label{tor2}
\end{equation}
Suppose $f$ is a solution of the first equation with parameter $B$. 
Then for any $c>0$, the function $cf$ is a solution
of (\ref{tor1}) with $B$ replaced by $cB$. At the same time, this rescaling 
does not affect (\ref{tor2}) in any way, since the ratio $B/f$ remains fixed.
However, the system behaves differently with respect to rescaling of the 
independent variable. Indeed, suppose $f$ satisfies both equations with 
parameters $B$, $\nu$, and $N$, 
then defining $g(x,y) = f(cx,cy)$, we have that $g$ 
satisfies (\ref{tor1}) and  (\ref{tor2}) with parameters
$Bc$, $\nu c^2$, and $N/c$, while its period is $1/c$ in both directions. 
This is consistent with the experimental fact that as a 
magnetic field normal to
the surface increases, vortices should eventually collide with one another.
(In physical reality they will at that point disappear together with the 
superconducting state).
This fact is also important mathematically, as it allows us to first obtain a 
solution of (\ref{tor1}) with, say, $\int_{T^2}\frac{B}{f}dV < 2\pi$, 
and then rescale the independent variable to satisfy (\ref{tor2}). 

It appears that the system (\ref{tor1})-(\ref{tor2}) does
not subdue itself to the standard techniques of variational calculus
or topological analysis. In particular, perturbative methods
do not apply to (\ref{tor1}), and no solutions arise as a result of 
bifurcation. In fact, in view of the theorem 
below and the results of numerical simulations,
solutions are objects very unlike the familiar vortex solutions
of nonlinear PDEs. To give some indication of the difficulties involved, 
consider the following.
The equation (\ref{tor1}) is the Euler-Lagrange equation for the Lagrangian
$L(f)= (\frac{1}{2}\int |\nabla f|^2 + B^2\int \ln(f))/(\int f^2)$. However, 
this functional is neither bounded below nor above. Indeed, let us note that 
$L(c_n)\rightarrow -\infty$ for constants $c_n \rightarrow 0$. On the other 
hand, let $f_n(x,y) = 1+\varepsilon + cos(2\pi n x)$. Since 
$\ln f_n(x,y) \geq \ln \varepsilon$, one easily checks that in this case
$L(f_n)\rightarrow \infty$. Therefore, the best we can expect is to
discover {\em local} extrema. Additional difficulty stems from the fact that
since (\ref{tor1}) always admits a trivial constant solution,  
we must devise a method of telling trivial and nontrivial solutions apart.

We will now consider a finite difference model of the system 
(\ref{tor1})-(\ref{tor2}). It is proven below that non-constant solutions 
of the discrete problem exist. 
The proof is independent of the number of points in the discretization ($n^2$) 
but relies on finite-dimensionality essentially,
and does not admit a direct generalization to the analytic case. 
However, all the universal parameters used in the proof,
like the $L_2$-norm of $f$ and $B$, are asymptoticly independent of $n$. 
Thus, we conjecture existence of the continuous domain solutions of 
(\ref{tor1}) that satisfy the equation a.e. in the classical sense 
and retain the particular vortex morphology.

It is convenient to introduce the following notation. 
$\int = \frac{1}{n^2}\sum\limits_i\sum\limits_j$,
$\int\limits_ o = \frac{1}{n^2}\sum\limits_{i\neq i_0}\sum\limits_{j\neq j_0}$,
where indices $i,j$ run through the discrete $n$-by-$n$ lattice.
Also, $\bigtriangleup$ denotes the common five-point periodic discrete Laplacian,
i.e.
$
\bigtriangleup f(\frac{i}{n},\frac{j}{n}) = n^2
\left(f(\frac{i+1}{n},\frac{j}{n})+f(\frac{i}{n},\frac{j+1}{n})
+f(\frac{i-1}{n},\frac{j}{n})+f(\frac{i}{n},\frac{j-1}{n})
-4f(\frac{i}{n},\frac{j}{n})\right)
$ 
and $\nabla = (\frac{\partial}{\partial x},\frac{\partial}{\partial y})$ 
is the simplest two-point periodic gradient, i.e. say
$
\frac{\partial}{\partial x}f(\frac{i}{n},\frac{j}{n}) = n
(f(\frac{i+1}{n},\frac{j}{n})-f(\frac{i}{n},\frac{j}{n})).
$
In particular, the discrete integration-by-parts formula holds, i.e. 
$-\int(\bigtriangleup f)g = \int (\nabla f, \nabla g)$. 
Consider the function 
\[ \Phi(f) = \frac{1}{2}\int |\nabla f|^2 + B^2\int \ln(f).
\]
Pick arbitrary real numbers $a,b,c$,
fix a point $(x_0,y_0)=(\frac{i_0}{n},\frac{j_0}{n})$, and a number 
\begin{equation}
m_n = \left(\frac{n^2}{b}-\frac{1}{c}\right)^{-1}.
\label{the_m}
\end{equation}
Two submanifolds in $R^{n^2}$, 
\begin{equation}
\label{D}
 D^n_{a,b,c} = \{f>0: \int f^2 = a^2, \int\frac{1}{f} \leq \frac{1}{b},
               \min f = f(x_0,y_0) = m_n \}
\end{equation}
and its boundary
\begin{equation}
\label{delD}
 \partial D^n_{a,b,c} = \{f>0: \int f^2 = a^2, \int\frac{1}{f} = \frac{1}{b},
               \min f = f(x_0,y_0) = m_n \},
\end{equation}
play a fundamental role in understanding the nature of critical points of $\Phi$.
Depending on the particular value of $a$, $b$, and $c$, the set
$D^n_{a,b,c}$ is either empty, an $(n^2-2)$-dimensional spherical disk, or 
it degenerates to a point. Consider the hyper-plane
$H_n = \{f:f(x_0,y_0) = m_n \}$.  $D^n_{a,b,c}$ is a 
spherical disk immersed in $H_n$ precisely when $V$, 
the point closest to the origin of an open submanifold given by 
$\int\limits_o\frac{1}{f} = \frac{1}{b} - \frac{1}{n^2m_n}$, is located inside 
the ball $\int\limits_o f^2 \leq a^2-\frac{m_n^2}{n^2}$. One easily finds 
$V(x,y) = const = (n^2-1)c$ for all 
$x\neq x_0, y\neq y_0$, and for it to be inside 
the ball, it is necessary and sufficient that $a$ and $c$ satisfy 
\begin{equation}
\frac{(n^2-1)^3}{n^2}c^2 < a^2 - \frac{m_n^2}{n^2}.
\label{a_and_c}
\end{equation}
Conversely, if condition (\ref{a_and_c}) holds then $D^n_{a,b,c}$ is 
a nonempty spherical disk. Formally, one needs to check in addition that $m_n$ 
is indeed a minimum of every function that satisfies the two other conditions
in (\ref{D}), but this is straightforward. 
At this point I would like to point out that in the description of local 
minima of $\Phi$ below, the parameter $a$ is physical, and with good faith can be regarded
as the $L_2$-norm of the critical point, whereas the parameters $b,c$ are auxiliary
and will converge to $0$ as the density of discretization $n$ increases to infinity. 
In particular, interpreting $\frac{1}{b}$ as an
approximate value of the integral of the reciprocal of the function where $\Phi$
attains its local minimum is erroneous since the function 
develops a singularity at $(x_0,y_0)$. The following theorem will be proven.

\begin{th} Fix constants $B$ and $a$ as above. 
For a certain choice of the constants $b=b(n,a)$, and $c=c(n,a)$, 
the function $f\rightarrow\Phi(f)$ assumes local relative minima in $D^n_{a,b,c}$.
In particular,  the corresponding critical point, 
say $f_0>0$, satisfies the finite difference version of (\ref{tor1}) everywhere except 
one point, i.e.
\begin{equation}
 -\bigtriangleup f_0(x,y) +\frac{B^2}{f_0(x,y)}=\nu f_0(x,y).\qquad \mbox{for all}\quad
   (x,y) \neq (x_0,y_0).
\label{tor_dscr}
\end{equation}
Moreover, if $B$ is sufficiently large, then $f_0$ is not
a constant function. 
\label{minima}
\end{th} 
{\em Proof.}  It is convenient to first treat $D^n_{a,b,c}$ formally, 
and check that condition (\ref{a_and_c}) is satisfied later.
Proceeding formally, let $N$ denote an outward normal vector to 
$\partial D^n_{a,b,c}$ inside its ambient sphere, i.e. $N$ is 
tangent to the $(n^2-2)$-dimensional sphere 
$S_a = \{f:\int f^2 = a^2,\quad f(x_0,y_0) = m_n\}$ 
and points away from the region $D^n_{a,b,c}$. 
The main task is to show that $N\Phi>0$. 
It will then follow from smoothness of $\Phi$ ($\nabla$ 
is a linear operator and $f\rightarrow ln(f)$ is smooth for $f>0$) that it
assumes a local minimum inside $D^n_{a,b,c}$. Let us choose for $N$
the vector field defined by
\begin{equation} 
N_f(x,y) = \left\{\begin{array}{ll}
\frac{1}{ba^2}f(x,y) -\frac{1}{f(x,y)^2} &\qquad \mbox{for}\quad   (x,y) \neq (x_0,y_0)\\
 0 &\qquad \mbox{otherwise},
          \end{array}\right .
\end{equation}
where $f\in\partial D^n_{a,b,c}$.
A direct calculation shows that
\[ N_f\Phi(f) =  \frac{1}{ba^2}
               \left(-\int\limits_ of\bigtriangleup f  + B^2(1-\frac{1}{n^2})\right)
              + \int\limits_ o\frac{1}{f^2}\bigtriangleup f - B^2\int\limits_ o\frac{1}{f^3}. 
\]
It remains to analyze terms one by one.
First, since a function in $D^n_{a,b,c}$ assumes its minimum at $(x_0, y_0)$, we obtain
\begin{equation}
 -\int\limits_ of\bigtriangleup f = \int|\nabla f|^2 
             +\frac{1}{n^2}\bigtriangleup f(x_0,y_0)m_n \geq \int|\nabla f|^2 \geq 0.
\label{first}
\end{equation}
Next, by definition of $D^n_{a,b,c}$ and (\ref{the_m}) we obtain
\[ \frac{1}{b} = \int\frac{1}{f} = \int\limits_ o\frac{1}{f}+ \frac{1}{n^2}\frac{1}{m_n} = 
                 \int\limits_ o\frac{1}{f}+ \frac{1}{b} - \frac{1}{n^2c},            
\]
so that $\sum\limits_{i\neq i_0}\sum\limits_{j\neq j_0}\frac{1}{f} = \frac{1}{c}$
and therefore
\begin{equation}
 f(x,y) > c \qquad \mbox{for}\quad  (x,y) \neq (x_0,y_0). 
\label{lower_bound}
\end{equation}
As an immediate application we obtain 
\begin{equation}
\int\limits_ o\frac{1}{f^3} \leq \frac{1}{c^3}(1-\frac{1}{n^2}).
\label{second}
\end{equation}
Finally, we obtain the following inequality
\begin{equation}
\begin{array}{llll}
&& \int\limits_ o\frac{1}{f^2}\bigtriangleup f\\  &=& 
\frac{1}{n^2}\sum\limits_{i\neq i_0}\sum\limits_{j\neq j_0}
\frac{n^2}{f(\frac{i}{n},\frac{j}{n})^2}\left(f(\frac{i+1}{n},\frac{j}{n})+f(\frac{i}{n},\frac{j+1}{n})
+f(\frac{i-1}{n},\frac{j}{n})+f(\frac{i}{n},\frac{j-1}{n})
-4f(\frac{i}{n},\frac{j}{n})\right) \\
&\geq& \sum\limits_{i\neq i_0}\sum\limits_{j\neq j_0}\frac{1}{f(\frac{i}{n},\frac{j}{n})^2}
(-4f(\frac{i}{n},\frac{j}{n})) = 
-4\sum\limits_{i\neq i_0}\sum\limits_{j\neq j_0}\frac{1}{f(\frac{i}{n},\frac{j}{n})} \\
&=& -4n^2(\frac{1}{b}-\frac{1}{n^2}\frac{1}{m_n}) = -\frac{4}{c}.
\end{array}
\label{third}
\end{equation}
Together, estimates (\ref{first}), (\ref{second}), (\ref{third}) yield
\begin{equation}
  N_f\Phi(f) \geq (1-\frac{1}{n^2})B^2(\frac{1}{ba^2}-\frac{1}{c^3}) - \frac{4}{c}.
\label{Nf_est}
\end{equation}
So far no assumption has been made about the constants. Now, in order to guarantee 
existence of local minima one needs to ensure that  $D^n_{a,b,c}$ is nonempty, and
that the outer derivative is positive. Both these requirements are met if we pick 
\[ c = c(n) = \frac{a}{2n^2},
\]
and
\[ b = b(n) = \frac{a}{16n^6}.
\]
With this choice of $c$, (\ref{a_and_c}) holds and if $f$ satisfying 
the first two conditions in (\ref{D}), assumed at some other
point a value at least as small as $m_n$, we would have 
$\int\limits_o\frac{1}{f} > \frac{2m_n^{-1}}{n^2} = \frac{2}{b}-\frac{1}{cn^2} > 
\frac{1}{b}$, which is a contradiction.
Thus $D^n_{a,b,c}$ is nonempty.
On the other hand, inequality (\ref{Nf_est}) becomes 
$ N_f\Phi(f) \geq \frac{8B^2}{a^3}(1-\frac{1}{n^2})n^6 - \frac{8}{a}n^2$ which implies    
$ N_f\Phi(f)>0$ for $n$ sufficiently large. Consequently, $\Phi$
assumes a local minimum inside $D^n_{a,b,c}$. Equation (\ref{tor_dscr}) is 
automatically satisfied because it expresses 
the fact that the component of the derivative of $\Phi$
which is tangent to the sphere $\int f^2 = a^2$
 vanishes in all directions except possibly 
the $(x_0,y_0)$-direction.

Still, the local minimum $f_0$ could {\em a priori} be 
a function constant everywhere, except at the discontinuity in $(x_0,y_0)$. 
This can be avoided by taking $B$ sufficiently large.
As $B$ increases, the solution which is constant 
except at $(x_0,y_0)$, say
\[
f_1(x,y)=\left\{\begin{array}{r@{\quad:\quad}l}
        \kappa & (x,y)\neq(x_0,y_0) \\
		m_n & \mbox{otherwise}
		\end{array}\right.
\]
becomes unstable, 
i.e. it corresponds to a saddle point on the graph of $\Phi$. 
Indeed, one checks directly that 
\[
  \frac{d^2}{d\varepsilon ^2}\Phi(f_1+\varepsilon\phi ) =
     \int|\nabla \phi|^2 - B^2\int \frac{\phi ^2}{f_1^2},   
\]
for $\phi$ tangent to the origin-centered sphere at $f_1$, 
so that $\int f_1\phi = 0$. Now consider $\phi$ to be a
non-constant eigenfunction of the discrete Laplacian on a torus 
and let $\lambda$ denote the corresponding eigenvalue. 
Shifting it if necessary, we can assume that 
$\phi(x_0,y_0)=0$, so that $\phi$ is orthogonal to $f_1$.
We have that $\int|\nabla \phi|^2 = \lambda\int\phi ^2$, and thus
$\frac{d^2}{d\varepsilon ^2}\Phi(f_1+\varepsilon\phi ) < 0$
only if $\frac{B^2}{\kappa ^2} > \lambda$. Thus, $f_1$ is not a local
minimum for $B$ sufficiently large.
Moreover, since $\int f_1^2 = a^2$, $\kappa$ converges to $a$ 
as $n$ increases. Naturally, one can make sure that $\lambda$ remains fixed
regardless of $n$ by always picking $\phi$ to be a discretization 
of the same trigonometric function. This shows that a choice of $B$ which
guarantees that $f_0$ is a nontrivial solution does not depend on the
discretization. $\Box$

\section{Closing remarks}

The proof above depends essentially on the fact that all functions 
and manifolds are discrete. Indeed, the constants $b=b(n), c=c(n)$ we have
used tend to infinity as $n\rightarrow\infty$, and some of the estimates
make no sense in the limit. However, it is important that the magnetic induction $B$
and the $L_2$-norm of $f$ do not depend on $n$. On the other hand, simulation shows
that the discrete solution $f_0$ retains its particular morphology independently of 
$n$, and is always subharmonic. In addition, multiplying (\ref{tor_dscr}) by 
$f\chi_{\{(x,y) \neq (x_0,y_0)\}}$ and inspecting the vicinity of the singularity
we easily obtain $\nu \sim \frac{1}{a^2}\left(B^2 +\int |\nabla f|^2\right)$,
which is also expected to converge. 
In summary, we have enough evidence to believe that the figure above shows 
a good approximation to a strong solution of the continuous version of equation
(\ref{tor1}) that would posess a vortex of Lipschitz regularity. 
It is also consistent with the one-dimensional case,
where solutions of the analogous nonlinear equation can be expressed in terms of 
a closed-form integral.

I point out for completeness that continuity of such a positive
solution $f_0$ guarantees continuity of $F_A = \star\frac{B}{f_0}$, and by
rescaling the
independent variables we can satisfy the condition that the cycle of $F_A$ 
be an integral multiple of $2\pi$. This is sufficient to solve $dA=F_A$ for $A$
on a compact surface, and the solution $A$ has sufficient regularity to retain its
geometric interpretation as a connection $1$-form. On the other hand, $A$ 
contains all the information necessary to derive the basic tenets of 
superconductive electronics, like the Josephson effect.

It is important to realize that none of the propositions stated above
necessarily rely on the lattice being a simple square lattice. 
Most likely, a hexagonal lattice setting would yield 
the same qualitative results. The only reason for using a square 
lattice is to avoid overwhelming numerical complexity in experiments,
as well as arithmetical nuisance in theory.

I introduced the system (\ref{syst0})-(\ref{syst2}) in 1993, 
and later investigated its properties in my thesis. 
Some of those early results are contained in \cite{sow1}. 
The geometry was first given a loose but essentially correct 
physical interpretation in \cite{sow2}.

This work is a departure from the bulk of current research
I conduct in collaboration with Professor R. R. Coifman
in the area of Computational Harmonic Analysis. 
Working with Raphy is a stimulating adventure, and I want to thank him 
here for enhancing my understanding of mathematics.
I also wish to thank my friend Fred Warner for proofreading the manuscript.

\end{document}